\documentclass[prd, 
showpacs,nofootinbib,preprintnumbers]{revtex4}
\usepackage{amsmath}
\usepackage{amssymb}
\usepackage{amsfonts}
\usepackage{graphicx}
\usepackage{dcolumn}
\newcommand{\be}{\begin{equation}}
\newcommand{\ee}{\end{equation}}
\newcommand{\bea}{\begin{eqnarray}}
\newcommand{\eea}{\end{eqnarray}}
\newcommand{\beaa}{\begin{eqnarray*}}
\newcommand{\eeaa}{\end{eqnarray*}}

\newcommand{\nn}{\nonumber \\}
\newcommand{\e}{\mathrm{e}}

\begin{document}

\title{Unifying inflation with dark energy in modified $F(R)$
Ho\v{r}ava-Lifshitz gravity}

\author{E.~Elizalde$\,^{(a)}$,
S.~Nojiri$\,^{(b,c)}$,
S.~D. Odintsov$\,^{(a,d)}$\footnote{Also at TSPU, Tomsk, Russia.
} and
D.~S\'{a}ez-G\'{o}mez$\,^{(a)}$
\\ \vspace*{2mm}
%
{\it $^{(a)}$Consejo Superior de Investigaciones Cient\'{\i}ficas,
ICE(CSIC-IEEC), Campus UAB Facultat de Ci\`encies, Torre C5-Parell-2a
pl, E-08193 Bellaterra (Barcelona) Spain
\\ \vspace*{1mm}
$^{(b)}$Department of Physics, Nagoya University, Nagoya 464-8602,
Japan \\ \vspace*{1mm}
$^{(c)}$Kobayashi-Maskawa Institute for the Origin of Particles
and the Universe,\footnote{
See, http://www.kmi.nagoya-u.ac.jp/index-e.html .
}
Nagoya University, Nagoya 464-8602, Japan
\\ \vspace*{1mm}
$^{(d)}$ICREA, Barcelona, Spain}
}

\affiliation{\ }

\begin{abstract}

We study FRW cosmology for a non-linear modified $F(R)$ Ho\v{r}ava-Lifshitz
gravity which has a viable convenient counterpart. A unified description of
early-time inflation and late-time acceleration is possible in this theory,
but the cosmological dynamic details are generically different from the ones
of the convenient viable $F(R)$ model. Remarkably, for some specific
choice of parameters they do coincide. The emergence of finite-time
future singularities is investigated in detail.
It is shown that these singularities can be cured by adding an extra,
higher-derivative term, which turns out to be qualitatively different when
compared with the corresponding one of the convenient $F(R)$ theory.

\end{abstract}

\pacs{98.80.-k,04.50.+h,11.10.Wx}

\maketitle


\section{Introduction \label{I}}

Current observational data clearly indicates that our universe has
undergone at least two periods of accelerated expansion: the early-time
inflation and the present late-time cosmic acceleration.
In spite of the existence of a number of (partially) successful
scenarios for the inflationary and dark energy epochs, the fundamental issue of
a unified description of the whole cosmic history scenario remains open.
One possibility to solve this problem, relying only on the presence of gravity
(and not on the introduction of additional cosmological fields), is
modified gravity (for a quick presentation, see \cite{faraoni}).
Indeed, this approach suggests a very natural unification of
early-time inflation and late-time cosmic acceleration, as a purely
gravitational alternative (see \cite{review} for a review of such unified
models of modified gravity). But, how general is this scenario?

The Ho\v{r}ava-Lifshitz quantum gravity \cite{Horava}
has been conjectured to be renormalizable in four dimensions, at the price of
explicitly breaking Lorentz invariance. Its generalization to an
$F(R)$-formulation, which seems to be also renormalizable in $3+1$ dimensions,
has been considered in Refs.~\cite{FRhorava,FRhorava2}, where the Hamiltonian
structure and FRW cosmology, in a power-law theory, have been investigated for
such modified $F(R)$ Ho\v{r}ava-Lifshitz gravity. It was also conjectured there
that it sustains, in principle, the possibility of a unified description of
early-time inflation and the dark energy epochs.

The purpose of the present work is to study a realistic non-linear $F(R)$ gravity
in the Ho\v{r}ava-Lifshitz formulation, with the aim to understand if such theory
is in fact directly able to predict in a natural way the unification of the two
acceleration eras, similarly as its done in the convenient version.
It has been shown in \cite{FRhorava} that, for a special choice of parameters, 
the FRW equations
do coincide with the ones for the related, convenient $F(R)$ gravity.
This means, in particular, that the cosmological history of such
Ho\v{r}ava-Lifshitz $F(R)$ gravity will be just the same as for its convenient
version (whereas black hole solutions are generically speaking different).
For the general version of the theory the situation turns out to be more complicated.
Nevertheless, the unification of inflation with dark energy is still possible
and all local tests can also be passed, as we will prove.

Note that in the Ho\v{r}ava-Lifshitz formulation of $F(R)$ gravity,
the Lorentz symmetry is explicitly broken and the restoration of the Lorentz
symmetry at the observed energy scale is the main problem in this formulation.
The obtained metrics in this paper are, however, FRW metrics which are almost Lorentzian
at the scale of galaxy or solar system. Then when the matter sector has a Lorentz symmetry
in the flat background, if the Newton law is reproduced, the violation of the Lorentz
symmetry could be difficult to be observed. The restoration of the Newton law is actively
investigated but in this paper, we just show a mechanism that the scalar field corresponding
to the so-called scalaron in the usual $F(R)$-gravity does not give an observable correction
to the Newton law.
Although there is a problem with the Lorentz symmetry breaking,  the Ho\v{r}ava-Lifshitz
$F(R)$ gravity has a much richer structure than standard $F(R)$ gravity.

The paper is organized as follows.
Next section briefly reviews the formulation of modified $F(R)$
Ho\v{r}ava-Lifshitz gravity and its corresponding FRW cosmology.
The reconstruction of the theory is presented in Sect.~\ref{III}.
It is demonstrated that different functional forms of the theory may
reproduce the same $\Lambda$CDM era, phantom (super)acceleration,
or any other given cosmology. Sect.~\ref{IV} is devoted to the
analysis of $F(R)$ Ho\v{r}ava-Lifshitz theories
whose convenient counterparts have been proposed as viable candidates for
inflation-dark energy unification. The unification is again possible but
with rather different physical properties. Moreover, corrections
to Newton's law are negligible in the models under consideration
(Sect.~\ref{V}). The emergence of finite-time future singularities
and their avoidance is discussed in Sect.~\ref{VI}.
In particular, it turns out that even when the viable modified gravity is
non-singular, its Ho\v{r}ava-Lifshitz counterpart may still remain a singular theory.
Some discussions and an outlook are provided in the last section.

\section{Modified $F(R)$ Ho\v{r}ava-Lifshitz gravity \label{II}}

In this section, modified Ho\v{r}ava-Lifshitz $F(R)$
gravity is briefly reviewed \cite{FRhorava,FRhorava2}. We start by writing a
general metric in the so-called ADM decomposition in a $3+1$ spacetime (for
more details see \cite{ADM}, \cite{gravitation} and references
therein),
\be
ds^2=-N^2 dt^2+g^{(3)}_{ij}(dx^i+N^idt)(dx^j+N^jdt)\, ,
\label{1.1}
\ee
where $i,j=1,2,3$, $N$ is the so-called lapse variable, and $N^i$ is
the shift $3$-vector. In standard general relativity (GR),
the Ricci scalar can be written in terms of this metric, and yields
\be
R=K_{ij}K^{ij}-K^2+R^{(3)}+2\nabla_{\mu}(n^{\mu}\nabla_{\nu}n^{\nu}-n^{\nu}
\nabla_{\nu}n^{\mu})\, ,
\label{1.2}
\ee
here $K=g^{ij}K_{ij}$, $K_{ij}$ is the extrinsic curvature, $R^{(3)}$
is the spatial scalar curvature, and $n^{\mu}$ a unit vector
perpendicular to a hypersurface of constant time. The extrinsic
curvature $K_{ij}$ is defined as
\be
K_{ij}=\frac{1}{2N}\left(\dot{g}_{ij}^{(3)}-\nabla_i^{(3)}N_j-\nabla_j^{(3)}
N_i\right)\, .
\label{1.3}
\ee

In the original model \cite{Horava}, the lapse variable $N$ is taken
to be just time-dependent, so that the projectability condition holds
and by using the foliation-preserving diffeomorphisms (\ref{1.7}),
it can be fixed to be $N=1$.
As pointed out in \cite{Blas:2009qj},  imposing the projectability
condition may cause problems with Newton's law in the Ho\v{r}ava gravity.
On the other hand, Hamiltonian analysis shows that the non-projectable
$F(R)$-model is inconsistent \cite{masud}.
For the non-projectable case, the Newton law
could be restored (while keeping stability) by the ``healthy''
extension of the original Ho\v{r}ava gravity of Ref.~\cite{Blas:2009qj}.

The action for standard $F(R)$ gravity can be written as
\be
S=\int d^4x\sqrt{g^{(3)}}N F(R)\, .
\label{1.4}
\ee
Gravity of Ref.~\cite{Horava} is assumed to have different scaling
properties of the space and time coordinates
\be
x^i \to b x^i\, , \quad t \to b^zt\, ,
\label{1.6}
\ee
where $z$ is a dynamical critical exponent that renders the theory
renormalizable for $z=3$ in $3+1$ spacetime dimensions \cite{Horava}
(For a proposal of covariant renormalizable gravity with dynamical
Lorentz symmetry breaking, see \cite{no}).
GR is recovered when $z=1$. The scaling properties (\ref{1.6}) render
the theory  invariant only under the so-called foliation-preserving
diffeomorphisms:
\be
\delta x^i = \zeta(x^i,t)\, , \quad \delta t=f(t)\, .
\label{1.7}
\ee
It has been pointed that, in the IR limit, the full diffeomorphisms
are recovered, although the mechanism for this transition is not
physically clear. The action considered here was introduced
in Ref.~\cite{FRhorava},
\be
S=\frac{1}{2\kappa^2}\int dtd^3x\sqrt{g^{(3)}}N F(\tilde{R})\, , \quad
\tilde{R}= K_{ij}K^{ij}-\lambda K^2 + R^{(3)}+
2\mu\nabla_{\mu}(n^{\mu}\nabla_{\nu}n^{\nu}-n^{\nu}\nabla_{\nu}n^{\mu})-
L^{(3)}(g_{ij}^{(3)})\, ,
\label{1.8}
\ee
where $\kappa$ is the dimensionless gravitational coupling, and where, two
new constants $\lambda$ and $\mu$ appear, which account for the violation
of the full diffeomorphism transformations.
A degenerate version of the above $F(R)$-theory with $\mu=0$ has been proposed and
studied in Ref.~\cite{kluson}.
Note that in the original Ho\v{r}ava gravity theory \cite{Horava},
the third term in the expression for $\tilde{R}$ can be omitted, as
it becomes a total derivative. The term $L^{(3)}(g_{ij}^{(3)})$ is
chosen to be \cite{Horava}
\be
L^{(3)}(g_{ij}^{(3)})=E^{ij}G_{ijkl}E^{kl}\, ,
\label{1.9}
\ee
where $G_{ijkl}$ is the inverse of the generalized De Witt metric, namely
\be
G^{ijkl}=\frac{1}{2}\left(g^{(3)\, ik}g^{(3)\, jl}+g^{(3)\, il}g^{(3)\, jk}\right)
 - \lambda g^{(3)\, ij} g^{(3)\, kl}\, .
\label{1.10}
\ee
Therefore we have
\be
\label{HLF7b0}
G_{ijkl} = \frac{1}{2} \left( g^{(3)}_{ik} g^{(3)}_{jl} + g^{(3)}_{il} g^{(3)}_{jk} \right)
 - \tilde{\lambda} g^{(3)}_{ij} g^{(3)}_{kl}\, ,\quad
\tilde{\lambda} = \frac{\lambda}{3\lambda - 1}\, .
\ee
Note that $G^{ijkl}$ is singular for $\lambda=1/3$ and therefore $G_{ijkl}$ exist
if $\lambda \neq 1/3$.

In Ref.~\cite{Horava}, the expression for $E_{ij}$ is constructed to
satisfy the ``detailed balance principle'' in order to restrict the
number of free parameters of the theory. This is defined through
variation of an action
\be
\sqrt{g^{(3)}}E^{ij}=\frac{\delta W[g^{(3)}_{kl}]}{\delta g^{(3)}_{ij}} \, ,
\label{1.11}
\ee
where the form of $W[g^{(3)}_{kl}]$ is given in Ref.~\cite{Horava2} for
$z=2$ and  $z=3$. Other forms for
$L^{(3)}(g_{ij}^{(3)})$ have been suggested that abandons the
detailed balance condition but still render the theory power-counting
renormalizable (see Ref.~\cite{FRhorava2}).

We are interested in the study of (accelerating) cosmological
solutions for the theory described by action (\ref{1.8}).
Spatially-flat FRW metric is assumed
\be
ds^2=-N^2dt^2+a^2(t)\sum_{i=1}^3 \left(dx^{i}\right)^2\, .
\label{1.14}
\ee
If we also assume the projectability condition,
$N$ can be taken to be just time-dependent and, by using the
foliation-preserving
diffeomorphisms (\ref{1.7}), it can be fixed to be unity, $N=1$.
When we do not assume the projectability condition, $N$ depends on both
the time and spatial coordinates, first. Then, just as an assumption of
the solution, $N$ is taken to be unity.

For the metric (\ref{1.14}), the scalar $\tilde{R}$ is
given by
\be
\tilde{R}=\frac{3(1-3\lambda
+6\mu)H^2}{N^2}+\frac{6\mu}{N}\frac{d}{dt}\left(\frac{H}{N}\right)\, .
\label{1.15}
\ee
For the action (\ref{1.8}), and assuming the FRW metric (\ref{1.15}),
the second FRW equation can be obtained by varying
the action with respect to the spatial metric $g_{ij}^{(3)}$, which
yields
\be
0=F(\tilde{R})-2(1-3\lambda+3\mu)\left(\dot{H}+3H^2\right)F'(\tilde{R})-2(1-
3\lambda)
\dot{\tilde{R}}F''(\tilde{R})+2\mu\left(\dot{\tilde{R}}^2F^{(3)}(\tilde{R})
+\ddot{\tilde{R}}F''(\tilde{R})\right)+\kappa^2p_m\, ,
\label{1.16}
\ee
here $\kappa^2=16\pi G$, $p_m$ is the pressure of a perfect fluid
that fills the Universe, and $N=1$. Note that this
equation becomes the usual second FRW equation for convenient $F(R)$
gravity (\ref{1.4}), by setting the constants $\lambda=\mu=1$.
When we assume the projectability condition,
variation over $N$ of the action (\ref{1.8}) yields the following
global constraint
\be
0=\int d^3x\left[F(\tilde{R})-6(1-3\lambda +3\mu)H^2-6\mu\dot{H}+6\mu
H \dot{\tilde{R}}F''(\tilde{R})-\kappa^2\rho_m\right]\, .
\label{1.17}
\ee
Now,  using the ordinary conservation equation for the matter fluid
$\dot{\rho}_m+3H(\rho_m+p_m)=0$, and integrating Eq.~(\ref{1.16}),
\be
0=F(\tilde{R})-6\left[(1-3\lambda
+3\mu)H^2+\mu\dot{H}\right]F'(\tilde{R})+6\mu H
\dot{\tilde{R}}F''(\tilde{R})-\kappa^2\rho_m-\frac{C}{a^3}\, ,
\label{1.18}
\ee
where $C$ is an integration constant, taken to be zero,
according to the constraint equation (\ref{1.17}).
If we do not assume the projectability condition, we can directly
obtain (\ref{1.18}), which corresponds to the first FRW equation, by
variation over $N$.
Hence, starting from a given $F(\tilde{R})$ function, and solving
Eqs.~(\ref{1.16}) and (\ref{1.17}), a cosmological solution
can be obtained.

\section{Reconstructing FRW cosmology in $F(R)$ Ho\v{r}ava-Lifshitz
gravity \label{III}}

To start, let us analyze the simple model $F(\tilde{R})=\tilde{R}$,
which cosmology was studied in \cite{cosm} (for a complete analysis of
cosmological perturbations, see \cite{misao}). In such a case, the FRW
equations look similar to GR,
\be
H^2=\frac{\kappa^2}{3(3\lambda-1)}\rho_m\, , \quad
\dot{H}=-\frac{\kappa^2}{2(3\lambda-1)}(\rho_m+p_m)\, ,
\label{2.1}
\ee
where, for $\lambda\rightarrow 1$, the standard FRW equations are
recovered. Note that the constant $\mu$ is now irrelevant because, as
pointed out above, the term in front of $\mu$ in (\ref{1.8}) becomes a
total derivative. For such
theory, one has to introduce a dark energy source as well as an
inflaton field, in order to reproduce the cosmic and inflationary
accelerated epochs, respectively. It is also important to note that,
for this case, the coupling constant is restricted to be
$\lambda>1/3$, otherwise Eqs.~(\ref{2.1}) become
inconsistent. It seems reasonable to think that, for the current epoch,
where $\tilde{R}$ has a small value, the IR limit of the theory is
satisfied $\lambda\sim1$, but for the inflationary epoch, when the
scalar curvature $\tilde{R}$ goes to infinity, $\lambda$ will take a
different value. It has been realized that, for $\lambda=1/3$, the
theory develops an anisotropic Weyl invariance (see \cite{Horava}),
and thus it takes an special role, although for the present model this
value is not allowed.

We now discuss some cosmological solutions of
$F(\tilde{R})$ Ho\v{r}ava-Lifshitz gravity. The first FRW equation,
given by (\ref{1.18}) with $C=0$, can be rewritten as a function of
the number of e-foldings $\eta=\ln\frac{a}{a_0}$, instead of the usual
time $t$. This technique has been developed in Ref.~\cite{FR1} for
convenient $F(R)$ gravity (for review of reconstruction method in
modified gravity, see \cite{rec}), where it was shown that any $F(R)$
theory can be reconstructed for a given cosmological solution. Here,
we extend such formalism to the Ho\v{r}ava-Lifshitz $F(R)$ gravity. Since
$\frac{d}{dt}=H\frac{d}{d\eta}$ and $\frac{d^2}{d\eta^2}
=H^2\frac{d^2}{d\eta^2}+H\frac{dH}{d\eta}\frac{d}{d\eta}$,
the first FRW equation (\ref{1.18}) is rewritten as
\be
0=F(\tilde{R})-6\left[(1-3\lambda+3\mu)H^2+\mu
HH'\right]\frac{dF(\tilde{R})}{d\tilde{R}}
+36\mu H^2\left[(1-3\lambda+6\mu)HH'+\mu H'^2+\mu H''H \right]
\frac{d^2F(\tilde{R})}{d^2\tilde{R}}-\kappa^2\rho_m\, ,
\label{D1}
\ee
where the primes denote derivatives with
respect to $\eta$. Thus, in this case there is no restriction on the
values of $\lambda$ or $\mu$.  By using the energy conservation
equation, and assuming a perfect fluid with equation of state (EoS)
$p_m=w_m\rho_m$, the energy density yields
\be
\rho_m=\rho_0 a^{-3(1+w_m)}=\rho_0a_0^{-3(1+w_m)}\e^{-3(1+w_m)\eta}\, .
\label{D2}
\ee
As the Hubble parameter can be written as a function of the number of
e-foldings, $H=H(\eta)$, the  scalar curvature in (\ref{1.15}) takes the
form
\be
\tilde{R}=3(1-3\lambda+6\mu)H^2+6\mu HH'\, ,
\label{D3}
\ee
which can be solved with respect to $\eta$ as $\eta=\eta(\tilde{R})$,
and  one gets an expression (\ref{D1}) that gives an equation on
$F(\tilde{R})$ with the variable $\tilde{R}$. This can be
simplified a bit by writing $G(\eta)=H^2$ instead of the Hubble parameter. In
such case, the differential equation (\ref{D1}) yields
\be
0=F(\tilde{R})-6\left[(1-3\lambda+3\mu)G+\frac{\mu}{2}G'\right]\frac{dF(\tilde{R})
}{d\tilde{R}}+18\mu\left[(1-3\lambda+6\mu)GG'+\mu GG''\right]
\frac{d^2F(\tilde{R})}{d^2\tilde{R}}-\kappa^2\rho_0a_0^{-3(1+w)}
\e^{-3(1+w)\eta}\, ,
\label{D4}
\ee
and the scalar curvature is now written as
$\tilde{R}=3(1-3\lambda+6\mu)G+3\mu
G'$.  Hence,
for a given cosmological solution $H^2=G(\eta)$, one can resolve
Eq.~(\ref{D4}), and the $F(\tilde{R})$ that reproduces such solution
is obtained.

As an example, we consider the Hubble parameter that reproduces the
$\Lambda$CDM epoch. It is expressed as
\be
H^2 =G(\eta)= H_0^2 + \frac{\kappa^2}{3}\rho_0 a^{-3} = H_0^2 +
\frac{\kappa^2}{3}\rho_0 a_0^{-3} \e^{-3\eta} \, .
\label{D5}
\ee
where $H_0$ and $\rho_0$ are constant. In General
Relativity, the terms on the rhs of Eq.~(\ref{D5}) correspond to an
effective cosmological constant $\Lambda=3H_0^2$ and to cold dark
matter with EoS parameter $w=0$. The corresponding
$F(\tilde{R})$ can be reconstructed by following the same steps as
described above. Using the expression for the scalar curvature
$\tilde{R}=3(1-3\lambda+6\mu)G+3\mu G'$,
the relation between $\tilde{R}$ and $\eta$ is obtained,
\be
\e^{-3\eta}=\frac{\tilde{R}
 -3(1-3\lambda+6\mu)H_0^2}{3k(1+3(\mu-\lambda))}\, ,
\label{D7}
\ee
where $k=\frac{\kappa^2}{3}\rho_0 a_0^{-3}$. Then, substituting
(\ref{D5}) and (\ref{D7}) into Eq.~(\ref{D4}), one gets the
differential expression
\bea
0 &=&
(1-3\lambda+3\mu)F(\tilde{R})-2\left(1-3\lambda+\frac{3}{2}\mu\right)
\tilde{R}
+9\mu(1-3\lambda)H^2_0\frac{dF(\tilde{R})}{d\tilde{R}} \nn
&& -6\mu(\tilde{R}-9\mu H^2_0)(\tilde{R}-3H^2_0(1-3\lambda+6\mu))
\frac{d^2F(\tilde{R})}{d^2\tilde{R}}-R-3(1-3\lambda+6\mu)H^2_0\, ,
\label{D8}
\eea
where, for simplicity, we have  considered a pressureless fluid $w=0$
in Eq.~(\ref{D4}). Performing the change of
variable $x=\frac{\tilde{R}-9\mu H^2_0}{3H^2_0(1+3(\mu-\lambda))}$,
the homogeneous part of
Eq.~(\ref{D8})  can be easily identified as an hypergeometric
differential equation
\be
0=x(1-x)\frac{d^2 F}{dx^2} + \left(\gamma - \left(\alpha + \beta +
1\right)x\right)\frac{dF}{dx} - \alpha \beta F\, ,
\label{D9}
\ee
with the set of parameters $(\alpha,\beta,\gamma)$ being given by
\be
\gamma=-\frac{1}{2}\, , \quad \alpha+\beta=
\frac{1-3\lambda-\frac{3}{2}\mu}{3\mu}\, , \quad
\alpha\beta=-\frac{1+3(\mu-\lambda)}{6\mu}\, .
\label{D10}
\ee
The complete solution of Eq.~(\ref{D9}) is a Gauss' hypergeometric function
plus a linear term and a cosmological constant coming from the particular
solution of Eq.~(\ref{D8}), namely
\be
F(\tilde{R}) = C_1 F(\alpha,\beta,\gamma;x) + C_2 x^{1-\gamma} F(\alpha -
\gamma + 1, \beta - \gamma +
1,2-\gamma;x)+\frac{1}{\kappa_1}\tilde{R}-2\Lambda\, .
\label{D11}
\ee
where $C_1$ and $C_2$ are constants, $\kappa_1=3\lambda-1$ and
$\Lambda=-\frac{3H_0^2(1-3\lambda+9\mu)}{2(1-3\lambda+3\mu)}$.
Note that for the exact cosmology (\ref{D5}), the classical $F(R)$
gravity was reconstructed and studied in Refs.~\cite{FR1}-\cite{FRLAmbdaCDM}.
In this case, the solution (\ref{D11}) behaves similarly
to the classical $F(R)$ theory, except that now the parameters of
the theory depend on $(\lambda,\mu)$, which are
allowed to vary as it was noted above.
We can conclude that the cosmic evolution
described by the Hubble parameter (\ref{D5}) is reproduced by this class of
theories.

One can also explore the solution (\ref{D5}) for a particular choice
on the parameters $\mu=\lambda-\frac{1}{3}$, which plays a special role as
it is shown below. In this case, the scalar $\tilde{R}$ turns out to be a
constant,
and Eq.~(\ref{D8}), in the presence of a pressureless fluid, has the
solution
\be
F(\tilde{R})=\frac{1}{\kappa_1}\tilde{R}-2\Lambda\, , \quad \text{with}
\quad \Lambda=\frac{3}{2}(3\lambda-1)H_0^2\, .
\label{DD11}
\ee
Hence, for this constraint on the parameters,  the only consistent solution
reduces to the Ho\v{r}ava linear theory with a cosmological constant.

As a further example, we consider the so-called phantom accelerating
expansion.
Currently, observational data do not totally exclude the  possibility that
the Universe could have already crossed the phantom divide, which means
that the effective EoS for dark energy would presently
be slightly less than $-1$. Such kind of system can be easily expressed in GR,
where the FRW equation reads $H^2=\frac{\kappa^2}{3}\rho_\mathrm{ph}$. Here
the subscript ``ph'' denotes the phantom nature of the fluid, which has
an EoS given by $p_\mathrm{ph}=w_\mathrm{ph}\rho_\mathrm{ph}$
with $w_\mathrm{ph}<-1$. By using
the energy conservation equation, the solution for the Hubble
parameter turns out to be
\be
H(t)=\frac{H_0}{t_s-t}\, ,
\label{D12}
\ee
where $H_0=-1/3(1+w_\mathrm{ph})$, and $t_s$ is the Rip time which
represents the time still remaining up to the Big Rip singularity.
As in the above example, one can
rewrite the Hubble parameter as a function of the number of
e-foldings; this yields
\be
G(\eta)=H^2(\eta)=H^2_0\e^{2\eta/H_0}\, .
\label{D13}
\ee
Then, by using the expression of the scalar curvature, the relation
between $\tilde{R}$ and $\eta$ is given by
\be
\e^{2\eta/H_0}=\frac{R}{H_0(AH_0+6\mu)}\, .
\label{D14}
\ee
By inserting (\ref{D13}) and (\ref{D14}) into the differential
equation (\ref{D4}), we get
\be
\tilde{R}^2\frac{d^2F(\tilde{R})}{d\tilde{R}^2}+k_1\tilde{R}
\frac{dF(\tilde{R})}{d\tilde{R}}+k_0F(\tilde{R})=0\, ,
\label{D15}
\ee
where
\be
k_1=-\frac{(AH_0+6\mu)((AH_0+3\mu))}{6\mu(AH_0+12\mu)}\, , \quad
k_0=\frac{(AH_0+6\mu)^2}{12\mu(AH_0+12\mu)} \, ,
\label{D16}
\ee
here we have neglected any kind of matter contribution for simplicity.
Eq.~(\ref{D15}) is an Euler equation, whose solution is well known
\be
F(R)=C_1R^{m_+}+C_2R^{m_-}\, , \quad \text{where} \quad
m_{\pm}=\frac{1-k_1\pm\sqrt{(k_1-1)^2-4k_0}}{2}\, .
\label{D17}
\ee
Such theory belongs to the class of models with positive and negative
powers of the curvature introduced in Ref.~\cite{NO}.
The existence of the negative power of curvature indicates that 
the flat Minkowski space is not
realized. If the coefficient $C_2$ is small enough, however, 
the difference from
the flat Minkowski space could not be observed. Moreover, 
one of the reasons to 
consider non-linear models of next section is related with 
the fact that Minkowski solution may be realized there.

Hence, a $F(\tilde{R})$ Ho\v{r}ava-Lifshitz gravity has been reconstructed that
reproduces the phantom dark epoch with no need of any exotic fluid.
In the same way, any given cosmology may be reconstructed.

\section{Unified inflation and dark energy in modified
Ho\v{r}ava-Lifshitz gravity \label{IV}}

Let us consider here some viable $F(\tilde{R})$ gravities which admit
the unification \cite{NO} of inflation with late-time acceleration.
In the convenient $F(R)$ theory, a number of viable models
(see \cite{FR2}, \cite{FR3}) which pass all local tests and are able to
unify the inflationary
and the current cosmic accelerated epochs have been proposed. Here we
extend this class of
models to the Ho\v{r}ava-Lifshitz gravity. We consider the action,
\be
F(\tilde{R})=\tilde{R}+f(\tilde{R})\, ,
\label{3.1}
\ee
where it is assumed that the term $f(\tilde{R})$ becomes important at
cosmological scales, while for scales compared with the Solar system one
the theory becomes linear on $\tilde{R}$. As an example, we consider
the following function \cite{FR2}
\be
f(\tilde{R})=\frac{\tilde{R}^n(\alpha\tilde{R}^n-\beta)}{1+\gamma
\tilde{R}^n}\, ,
\label{3.2}
\ee
where $(\alpha, \beta, \gamma)$ are constants and $n>1$. This theory
reproduces the inflationary and cosmic acceleration epochs in convenient
$F(R)$ gravity (see Ref.~\cite{FR2}), which  is also the case  in the
present theory, as will be shown. During inflation, it is assumed
that the curvature scalar tends to infinity. In this case the model
(\ref{3.1}), with (\ref{3.2}), behaves as
\be
\lim_{\tilde{R}\rightarrow\infty}F(\tilde{R})=\alpha\tilde{R}^n\, .
\label{3.3}
\ee
Then, by solving the FRW equation (\ref{1.18}), this kind of function
yields a power-law solution of the type
\be
H(t)=\frac{h_1}{t}\, , \quad \text{where} \quad
h_1=\frac{2\mu(n-1)(2n-1)}{1-3\lambda+6\mu-2n(1-3\lambda+3\mu)}\, .
\label{3.4}
\ee
This solution produces acceleration during the inflationary epoch if
the parameters of the theory are properly defined. The acceleration
parameter is given by $\frac{\ddot{a}}{a}=h_1(h_1-1)/t^2$, thus, for
$h_1>1$ the inflationary epoch is well reproduced by the model
(\ref{3.2}). On the other hand, the function (\ref{3.2}) has a
minimum at $\tilde{R}_0$, given by
\be
\tilde{R}_0 \sim \left( \frac{\beta}{\alpha\gamma}\right)^{1/4}\, ,
\qquad f'(\tilde{R})=0\, , \qquad  f(\tilde{R})= -2\Lambda\sim
-\frac{\beta}{\gamma}\, ,
\label{3.5}
\ee
where we have imposed the condition $\beta\gamma/\alpha\gg 1$. Then, at
the current epoch the scalar curvature acquires a small value which
can be fixed to coincide with the minimum (\ref{3.5}), such that the
FRW equations (\ref{1.16}) and (\ref{1.18}) yield
\be
H^2=\frac{\kappa^2}{3(3\lambda-1)}\rho_m+\frac{2\Lambda}{3(3\lambda-1)}\,
\quad \dot{H}=-\kappa^2\frac{\rho_m+p_m}{3\lambda-1}\, ,
\label{3.6}
\ee
which look very similar to the standard FRW equations in General 
Relativity, except for the parameter $\lambda$. 
For first consideration of FRW eqs. in Ho\v{r}ava-Lifshitz gravity based on General Relativity 
see ref. \cite{Kiritsis:2009sh}.
As has been pointed out, at the current
epoch the scalar $\tilde{R}$ is  small, so the theory is
in the IR limit where the parameter $\lambda\sim1$, and the equations
approach the usual ones for $F(R)$ gravity. Hence, the FRW
equations (\ref{3.6}) reproduce the behavior of the well known
$\Lambda$CDM model with no need to introduce a dark energy fluid
to explain the current universe acceleration.

As another example of the models described by (\ref{3.1}), we
can considered the function \cite{FR3, HS},
\be
f(\tilde{R})=-\frac{(\tilde{R}-\tilde{R}_0)^{2n+1}
+\tilde{R}_0^{2n+1}}{f_0+f_1\left[(\tilde{R}-\tilde{R}_0)^{2n+1}
+\tilde{R}_0^{2n+1}\right]}=-\frac{1}{f_1}+\frac{f_0/f_1}{f_0
+f_1\left[(\tilde{R}-\tilde{R}_0)^{2n+1}+\tilde{R}_0^{2n+
1}\right]}\, .
\label{3.7}
\ee
This function could also serve for the unification of inflation and
cosmic acceleration but, in this case, when one takes the limit
$\tilde{R}\rightarrow\infty$, one gets
\be
\lim_{\tilde{R}\rightarrow\infty}F(\tilde{R})=\tilde{R}-2\Lambda_i\, ,
\quad \text{where} \quad \Lambda_i=1/2f_1\, ,
\label{3.8}
\ee
where the subscript $i$ denotes that we are in the inflationary
epoch. By inserting this into Eqs.~(\ref{1.16}) and (\ref{1.18}), the
FRW equations take the same form as in (\ref{3.6}). Then, for the
function (\ref{3.7}) the inflationary epoch is produced by an
effective cosmological constant, which implies that the parameter
$\lambda>1/3$, or the equations themselves will present
inconsistencies, as it was discussed in the above section. For the
current epoch, it is easy to see that the function (\ref{3.7})
exhibits a minimum for $\tilde{R}=\tilde{R}_0$, which implies, as in
the model above, an effective cosmological constant for late time
that can produce the cosmic acceleration. The emergence of matter dominance
before the dark energy epoch can be exhibited,
in analogy with the case of the convenient theory. Hence, we have
shown that the model (\ref{3.7}) also unifies the cosmic expansion history,
although with different properties during the inflationary epoch as
compared with the model (\ref{3.2}). This could be very important
for the precise study of the evolution of the parameters of the theory.

It is also interesting to explore the de Sitter solutions allowed by
the theory (\ref{1.8}). By taking $H(t)=H_0$, the FRW equation
(\ref{1.18}), in absence of any kind of matter and with $C=0$,
reduces to
\be
0=F(\tilde{R}_0)-6H^2_0(1-3\lambda+3\mu)F'(\tilde{R}_0)\, ,
\label{3.9}
\ee
which reduces to an algebraic equation that, for an specific  model,
can be solved yielding the possible de Sitter points allowed by the
theory. As an example, let us consider the model (\ref{3.2}), where
Eq.~(\ref{3.9}) takes the form
\be
\tilde{R}_0+\frac{\tilde{R}_0^n(\alpha\tilde{R}_0^n-\beta)}{1
+\gamma\tilde{R}_0^n}
+\frac{6H_0^2(-1+3\lambda-3\mu)\left[1+n\alpha\gamma\tilde{R}_0^{3n-1}
+\tilde{R}_0^{n-1}(2\gamma\tilde{R}_0-n\beta)+\tilde{R}_0^{2n-1}(\gamma^2\tilde{R}_0
+2n\alpha)\right]}{(1+\gamma\tilde{R}_0^n)^2}=0\, .
\label{3.10}
\ee
Here $\tilde{R}_0=3(1-3\lambda+6\mu)H^2_0$. By specifying the free
parameters of the theory, one can solve Eq.~(\ref{3.10}), which
yields several de Sitter points, as the one studied above. They can
be used to explain the coincidence problem, with the argument that
the present will not be the only late-time accelerated epoch
experienced by our Universe. In standard $F(R)$, it was found for
this same model that it contains at least two de Sitter points along
the cosmic history (see \cite{FR4}). In the same way, the second
model studied here (\ref{3.7}), provides several de Sitter points in
the course of the cosmic history. Note that when $\mu=\lambda-\frac{1}{3}$,
Eq.~(\ref{3.9}) turns out to be much more simple, it reduces
to $F(\tilde{R}_0)=0$, where the de Sitter points are the roots.
For example, for (\ref{3.10}) we have
$\tilde{R}_0(1+\gamma\tilde{R}_0^n)+\tilde{R}_0^n(\alpha\tilde{R}_0^n-\beta)
=0$,
where the number of positive roots (de Sitter points) depends
on the free parameters of the theory.

Summing up, it has been here shown that, also in $F(\tilde{R})$
Ho\v{r}ava-Lifshitz gravity, the so-called viable models, as
(\ref{3.2}) or (\ref{3.7}), can in fact reproduce the whole
cosmological history of the universe, with no need to involve any
extra fields or a cosmological constant.
We do not, of course, believe the complicated model (\ref{3.2}) or (\ref{3.7})
could be a final model. These models, at best, could be low energy effective theories
coming from a more beautiful final theory. Even in the usual quantum field theories,
the corresponding low energy effective theories are rather complicated due to the terms
induced by the quantum effects. However, our main motivation to consider precisely 
them is related with the fact that they pass the known local tests as usual $F(R)$ theories.

\section{Newton law corrections in $F(\tilde{R})$ gravity \label{V}}

As is well-known, modified gravity may lead to
violations of  local tests. We explore in this
section how to avoid these violations of Newton's law. It is  known that
$F(\tilde{R})$ theories include scalar particle. This scalar field could
give rise
to a fifth force and to variations of the Newton law, which can be
avoided by a kind of the so-called chameleon mechanism \cite{khoury}.
Note that scalars with time-dependent mass were also considered
in \cite{Mota:2003tc}.

In the original Ho\v{r}ava gravity, the projectability
condition may cause problems with the Newton law \cite{Blas:2009qj} but the
model without
the projectability condition could be inconsistent for the
$F(R)$-model \cite{masud}. The Newton law in the Ho\v{r}ava gravity
may be restored by the ``healthy''
extension \cite{Blas:2009qj}.
In this section, we do not discuss the gravity sector corresponding to
the Ho\v{r}ava gravity but we show
that the scalar mode, which also appears in the usual $F(R)$ gravity,
can decouple from gravity and matter, and then the scalar mode does not give
a measurable correction to Newton's law.

To show this, we consider a function of the type (\ref{3.1}), and
rewrite action (\ref{1.8}) as
\be
S=\int
dtd^3x\sqrt{g^{(3)}}N\left[(1+f'(A))(\tilde{R}-A)+A+f(A)\right]\, ,
\label{4.1}
\ee
where $A$ is an auxiliary scalar field. It is easy to see that
variation of  action (\ref{4.1}) over $A$ gives $A=\tilde{R}$.
Performing the conformal transformation
$g^{(3)}_{ij}=\e^{-\phi}\tilde{g}^{(3)}_{ij}$, with
$\phi=\frac{2}{3}\ln(1+f'(A))$, action (\ref{4.1}) yields
\be
S=\int dtd^3x\sqrt{g^{(3)}}\left[\tilde{K}_{ij}\tilde{K}^{ij}-\lambda\tilde{
K}^2+\left(-\frac{1}{2}+\frac{3}{2}\lambda-\frac{3}{2}\mu\right)
\dot{\tilde{g}}^{ij(3)}\tilde{g}_{ij}^{(3)}\dot{\phi}
+\left(\frac{3}{4}-\frac{9}{4}\lambda+\frac{9}{2}\mu\right)\dot{\phi}^2
 -V(\phi)+\tilde{L}(\tilde{g}^{(3)},\phi)\right]\, ,
\label{4.2}
\ee
where
$\tilde{K}_{ij}$ is the extrinsic curvature given by $\tilde{g}^{(3)}_{ij}$
as in (\ref{1.3}), $\tilde K = \tilde{g}^{(3)\,ij} \tilde{K}_{ij}$,
$\tilde{L}(\tilde{g}^{(3)},\phi)$ is the conformally transformed
term in (\ref{1.9}), and the scalar potential is
\be
V(\phi)=\frac{A(\phi)f'(A(\phi))-f(A(\phi))}{1+f'(A(\phi))}\, .
\label{4.3}
\ee
Note that, differently from the convenient $F(R)$, in action
(\ref{4.2}) there is a coupling term between the scalar field $\phi$
and the spatial metric $\tilde{g}_{ij}^{(3)}$, which can thus be
dropped, by imposing the following condition on the parameters
\cite{FRhorava2}:
\be
\mu=\lambda-\frac{1}{3}\, .
\label{4.4}
\ee
This condition also renders the theory power-counting renormalizable,
for the same $z$ as in the original Ho\v{r}ava model.

Let us now investigate the term (\ref{1.9}). As already pointed out,
for local scales, where the scalar curvature is assumed to be very
small, the theory enters the IR limit, where such  term could be
written as a spatial curvature,
\be
L^{(3)}(g^{(3)},\phi)\sim R^{(3)}\, .
\label{4.5}
\ee
Then, corrections to the Newton law will come from the coupling that
now appears between the scalar field and matter, which makes a test
particle to deviate from its geodesic path, unless the mass of the
scalar field is large enough (since then the effect could be very small).
The precise value can be calculated from
\be
m^2_{\phi}=\frac{1}{2}\frac{d^2V(\phi)}{d\phi^2}
=\frac{1+f'(A)}{f''(A)}-\frac{A+f(A)}{1+f'(A)}\, .
\label{4.6}
\ee
In view of that, we can now analyze the models studied in the last section. We are
interested to see the behavior at local scales, as on Earth, where
the scalar curvature is around $A=\tilde{R}\sim 10^{-50} \mathrm{eV}^2$, or
in the solar system, where $A=\tilde{R}\sim 10^{-61} \mathrm{eV}^2$. The
function
(\ref{3.2}) and its derivatives can be approximated around these
points as
\be
f(\tilde{R})\sim-\frac{\beta}{\gamma}\, , \quad f'(\tilde{R})\sim
\frac{n\alpha}{\gamma}R^{n-1}\, , \quad
f''(\tilde{R})\sim\frac{n(n-1)\alpha}{\gamma}R^{n-2}\, .
\label{4.7}
\ee
Then, the scalar mass for the model (\ref{3.2}) is  given
approximately by the expression
\be
m^2_{\phi}\sim\frac{\gamma\tilde{R}^{2-n}}{n(n-1)\alpha}\, ,
\label{4.8}
\ee
which  becomes $m^2_{\phi}\sim10^{50n-100}\mathrm{eV}^2$ on Earth and
$m^2_{\phi}\sim10^{61n-122}$ in the Solar System. We thus see that,
for $n>2$, the scalar mass would be sufficiently large in order to
avoid corrections to the Newton law. Even for the limiting case
$n=2$, the parameters $\gamma/\alpha$ can be chosen to be large
enough so that any violation of the local tests is avoided.

For the model (\ref{3.7}), the situation is quite similar. For
simplicity, we impose the following condition
\be
f_0\ll f_1\left[(\tilde{R}-\tilde{R}_0)^{2n+1}+
\tilde{R}_0^{2n+1}\right]\sim f_1\tilde{R}^{2n+1}\, .
\label{4.9}
\ee
Then, function (\ref{3.7}) and its derivatives can be written,
for small values of the curvature, as
\be
f(\tilde{R})\sim -\frac{1}{f_1}+\frac{f_0}{f_1^2R^{2n+1}}, \quad
f'(\tilde{R})\sim-\frac{(2n+1)f_0}{f_1^2\tilde{R}^{2(n+1)}}, \quad
f''(\tilde{R})\sim\frac{2(2n+1)(2n+2)f_0}{f_1^2\tilde{R}^{2n+3}}\, .
\label{4.10}
\ee
Using now the expression for the scalar mass (\ref{4.6}), this yields
\be
m^2_{\phi}\sim\frac{f_1^2\tilde{R}^{2n+3}}{2(2n+1)(n+1)f_0}
+\frac{1}{f_1}\, .
\label{4.11}
\ee
Hence, as $\tilde{R}$ is very small at solar or Earth scales
($\sim10^{-50\sim-61}\mathrm{eV}^2$), and
$\Lambda_i=\frac{1}{2f_1}\sim10^{20\sim38}\mathrm{eV}^2$ is the effective
cosmological constant at inflation, which is much larger than the
other term, it turns out that the second term on the rhs of
(\ref{4.11}) will dominate, and the scalar mass will take a value such as
$m^2_{\phi}\sim\frac{1}{f_1}\sim10^{20\sim38}\mathrm{eV}^2$, which is large
enough as compared with the scalar curvature. As a consequence there
is no observable correction to Newton's law.

We have thus shown, in all detail, that the viable models (\ref{3.2}) and
(\ref{3.7}) do not introduce any observable correction to the Newton
law at small scales. This strongly supports the choice of $F(\tilde{R})$
gravity as a realistic candidate for the unified description of the
cosmological history.

\section{Finite-time future singularities in  $F(\tilde{R})$ gravity
\label{VI}}

It is a well-known that a good number of effective
phantom/quintessence-like dark energy models end their evolution
at a finite-time future singularity. In the current section, we
study the possible future evolution of the viable $F(R)$ Ho\v{r}ava-Lifshitz
gravity considered above. It has been already proven \cite{FRhorava2} that power-law
$F(R)$ Ho\v{r}ava-Lifshitz gravities may lead, in its evolution, to a finite-time singularity.
In order to properly define the type of future singularities, let us rewrite the FRW
Eqs.~(\ref{1.16}) and (\ref{1.18}) in the following way
\be
3H^2=\frac{\kappa^2}{3\lambda-1}\rho_\mathrm{eff}\, , \quad
-3H^2-2\dot{H}=\frac{\kappa^2}{3\lambda-1}p_\mathrm{eff}\, ,
\label{5.1}
\ee
where
\bea
&&
\rho_\mathrm{eff}=\frac{1}{\kappa^2F'(\tilde{R})}\left[-F(\tilde{R})
+3(1-3\lambda
+9\mu)H^2 F'(\tilde{R})-6\mu\dot{H}F'(\tilde{R})-6\mu H
\dot{\tilde{R}}F''(\tilde{R})+\kappa^2\rho_m\right]\, , \nn
&&
p_\mathrm{eff}=\frac{1}{\kappa^2F'(\tilde{R})}
\left[F(\tilde{R})-(6\mu\dot{H}
+3(1-3\lambda+9\mu)H^2)F'(\tilde{R})-2(1-3\lambda)
\dot{\tilde{R}}F''(\tilde{R}) \right. \nn
&& \left. \qquad \quad +2\mu\left[\dot{\tilde{R}}^2F^{(3)}(\tilde{R})
+\ddot{\tilde{R}}F''(\tilde{R})\right] +\kappa^2p_m\right]\, .
\label{5.2}
\eea
Then, using the expressions for the effective energy and pressure
densities just defined, the list of future singularities, as classified in
Ref.~\cite{Singularities}, can be extended to $F(\tilde{R})$ gravity as follows:
\begin{itemize}
\item Type I (``Big Rip''): For $t\rightarrow t_s$, $a\rightarrow
\infty$ and $\rho_\mathrm{eff}\rightarrow \infty$, $|p|\rightarrow \infty$.
\item Type II (``Sudden''): For $t\rightarrow t_s$, $a\rightarrow
a_s$ and $\rho_\mathrm{eff}\rightarrow \rho_s$, $|p_\mathrm{eff}|\rightarrow
\infty$.
\item Type III: For $t\rightarrow t_s$, $a\rightarrow a_s$
and $\rho_\mathrm{eff}\rightarrow \infty$,
$|p_\mathrm{eff}|\rightarrow \infty$.
\item Type IV: For $t\rightarrow t_s$, $a\rightarrow a_s$ and
$\rho_\mathrm{eff}\rightarrow \rho_s$, $p_\mathrm{eff} \rightarrow p_s$
but higher derivatives of Hubble parameter diverge.
\end{itemize}

To illustrate the possibility of future singularities in viable
$F(\tilde{R})$ gravity, we
explore the model (\ref{3.2}), which has been studied in
Ref.~\cite{FrSing} for the convenient $F(R)$ case, where it was shown
that this model is non-singular, in the
particular case $n=2$. Also, it is known that for most models of $F(R)$
gravity, the future singularity  can be cured by adding a term
proportional to $R^2$ (see Ref.~\cite{R2Sing,bamba}) or a non-singular
modified gravity action \cite{FrSing}. However, this is not
possible to do in the Ho\v{r}ava-Lifshitz gravity where,
even for the simple model
(\ref{3.2}) with  $n=2$, a singularity can occur unless some
restrictive conditions on the parameters are imposed.

In order to study the possible singularities that may occur from the above list,
we consider a Hubble parameter close to one of such singularities,
given by the following expression
\be
H(t)=\frac{h_0}{(t_s-t)^{q}}\, ,
\label{5.3}
\ee
where $h_0$ and $q$ are constant. Depending on the value of $q$, the
Hubble parameter (\ref{5.3}) gives rise to a particular type of
singularity. Thus,  for $q\geq 1$,
it gives a Big Rip singularity, for $-1<q<0$ a Sudden Singularity,
for $0<q<1$ a Type III singularity and for $q<-1$, it will produce a
Type IV singularity. Using the expression (\ref{1.15}) with
(\ref{5.3}), the scalar curvature can be approximated, depending on
the value of $q$, as
\be
\label{5.4}
R\sim \left\{ \begin{array}{lll}
\frac{3(1-3\lambda+6\mu)h_0^2}{(t_s-t)^{2q}} \quad & \mbox{for}
\quad & q>1 \\
\frac{3(1-3\lambda+6\mu)h_0^2+6\mu q h_0}{(t_s-t)^{2}} \quad &
\mbox{for} \quad & q=1 \\
\frac{6\mu q h_0}{(t_s-t)^{q+1}} \quad & \mbox{for} \quad & q<1
\end{array} \right. \, .
\ee
The model (\ref{3.2}) which, as has been shown, unifies
the inflationary and the dark energy epochs, makes the scalar
curvature grow with time so that, close to a possible future
singularity,  the model can be approximated as
$F(\tilde{R})\sim\tilde{R}^n$, where  $n>1$. For
the case $q>1$, it is possible to show, from the FRW
Eqs.~(\ref{1.16}) and (\ref{1.18}), that the solution (\ref{5.3}) is
allowed for this model just in some special cases: (i) For $n=3/2$
and $\mu=2\lambda-\frac{2}{3}$, which contradicts the decoupling
condition (\ref{4.4}) and the Newtonian corrections (where it was found
that $n>2$). \
(ii) When  $\mu=0$ and $\lambda=1/3$, which holds (\ref{4.4})
and fixes the values of the parameters, in contradiction with the fact that
fluctuations are allowed for them.

For $q=1$, the Hubble parameter (\ref{5.3}) is a natural solution for
this model,  yielding
\be
H(t)=\frac{h_0}{t_s-t} \quad \mbox{with} \quad
h_0=\frac{2\mu(n-1)(2n-1)}{-1+3\lambda-6\mu+2n(1-3\lambda+3\mu)}\, .
\label{5.5}
\ee
This model allows for the possibility of occurrence of a Big Rip
singularity \cite{rip}, unless the parameters of the theory are fixed. As
pointed out in Ref.~\cite{FrSing}, for the case $n=2$
the Big Rip singularity can be avoided in standard $F(R)$ gravity,
which can be easily seen by choosing $\lambda=\mu=1$ in the solution
(\ref{5.5}). Nevertheless, in $F(\tilde{R})$ gravity, in order to
avoid such singularity, the power $n$ of the model has to be fixed to
the value
\be
n=\frac{1}{2}+\frac{3\mu}{2-6\lambda+6\mu}\, ,
\label{5.6}
\ee
which can be interpreted as another constraint on the parameters of
the theory (although, when the decoupling condition (\ref{4.4}) is satisfied,
no constraints can be imposed on $n$). In addition, note that $h_0=-h_1$
in (\ref{3.4}), where
we imposed $h_1>1$ with the aim to reproduce the inflationary epoch,
so that $h_0$ would be negative and the solution (\ref{5.5}) would
correspond to the Big Bang singularity. Thus, no future doomsday
will take place.

For the last case, when $q<1$, we  find that the Hubble
parameter (\ref{5.3}) can be a consistent solution when Eq.~(\ref{4.4})
is satisfied
and $q=(n(1-n)-1)/n(n-2)$, although if we impose $n\geq 2$, as it was found
above, $q<-1$,
which implies a future singularity of the type IV. The exception here is
when $n=2$, that avoids the occurrence of any type of singularity when $q<1$.
Nevertheless, even in the case that, for any reason, some kind of
future singularity would be allowed in the model (\ref{3.2}), one must
also take into account possible quantum gravity effects, which may
probably become important when one is close to the singularity. They
have been shown, in phantom models, to prevent quite naturally the
occurrence of the future singularity \cite{eli}.

In summary, we have here proven that the theory (\ref{3.2}) can
actually be free of future singularities, and thus that it can make a good
candidate for the unification of the cosmic history. Note that, from
the corrections to the Newton law studied in the previous section, as
well as from imposing avoidance of future singularities, we can
fix some of the parameters of the theory. The other constant
parameters that appear in (\ref{3.2})---which expresses the algebraic
relation between the powers of the scalar curvature---could be fixed
by comparing the cosmic evolution with the observed data, quite in
the same way as has been successfully done for standard $F(R)$ gravity in
Ref.~\cite{FR4}.

\section{Discussion \label{VII}}

In summary, we have here investigated the FRW cosmology of a non-linear modified
Ho\v{r}ava-Lifshitz $F(R)$ gravity theory which has a viable convenient
counterpart. As shown in \cite{FRhorava}, for a special choice of the parameters,
the FRW equations are just the same in both theories, and that the cosmic
history of the first literally coincides with the one for the viable $F(R)$
gravity.
For a more general version of the theory, the unified description of the
early-time inflation and late-time acceleration is proven to be possible
too; however, the details of the cosmological dynamics are here different.
Moreover, corrections to Newton's law are negligible for an extensive region of
the parameter space. We have demonstrated the emergence of possible finite-time future
singularities, and their avoidance, by adding extra higher-derivative terms
which turn out to be qualitatively different, as compare with conventional
$F(R)$ cosmology.

Using the approach of Refs.~\cite{FRhorava, FRhorava2}, one can construct more
complicated generalizations of modified gravity with anisotropic scaling
properties. For instance, one can obtain non-local or modified Gauss-Bonnet
Ho\v{r}ava-Lifshitz gravities. Technically, this is of course a more involved
task, as compared with the $F(R)$ case. The corresponding cosmology can in
principle be studied in the same way as in the present work, with expected
qualitatively similar results, owing to the fact that the
convenient cosmological model has been already investigated, in those cases.
It is foreseen that the study of such theories will help us also in the resolution of
the some problems for the theories with broken Lorentz symmetry. For instance,
it is quite possible that a natural scenario of dynamical Lorentz symmetry
breaking, which would reduce gravity to the Ho\v{r}ava limit, may be found in this way,
which would be certainly interesting.


\section*{Acknowledgements}

This work has been supported in part by MICINN
(Spain), project FIS2006-02842, by AGAUR (Generalitat
de Ca\-ta\-lu\-nya), contract 2009SGR-994, by Global COE Program of Nagoya
University (G07) provided by the Ministry of Education,Culture, Sports,
Science
\& Technology of Japan, and by the JSPS Grant-in-Aid for Scientific Research
(S) \# 22224003.
DSG acknowledges an FPI grant from MICINN.


\end{document}